\documentclass[jkps,preprint,fleqn,showpacs,showkeys,nofootinbib]{revtex4}
\usepackage{graphicx}
\usepackage{amssymb}
\usepackage{amsmath}
\usepackage{bm}

\usepackage{amsfonts,amsmath,amssymb}
\usepackage{enumerate}
\usepackage{hyperref}
\usepackage{enumitem}

\newcommand{\be}{\begin{equation}}
\newcommand{\ee}{\end{equation}}

\newcommand{\bea}{\begin{eqnarray}}
\newcommand{\eea}{\end{eqnarray}}

\newcommand{\bes}{\begin{subequations}}
\newcommand{\ees}{\end{subequations}}

\newcommand{\nn}{\nonumber}

\begin{document}
\pagenumbering{gobble}
\title[]{On the U(1)$^2$-invariant sector of dyonic maximal supergravity}
\author{Hyojoong \surname{Kim}}
\email{h.kim@khu.ac.kr}
\author{Nakwoo \surname{Kim}}
\email{nkim@khu.ac.kr}
\affiliation{Department of Physics 
and Research Institute of Basic Science, \\ Kyung Hee University, Seoul 02447, Korea}
\author{Minwoo \surname{Suh}}
\email{minwoosuh1@gmail.com}
\affiliation{Department of Physics, Kyungpook National University, Daegu 41566, Korea}
\begin{abstract}
Recently a new maximally supersymmetric, dyonically gauged supergravity in four-dimenions has been constructed. This theory admits several supersymmetric AdS solutions, and a Chern-Simons-matter dual theory has been proposed for a solution with an unbroken SU(3) symmetry. The gravity side and the field theory side of the free energy computation show nice agreements at superconformal point. With a view to a study of non-conformal deformations, in this paper we study a consistent truncation of the full theory where SU(3) is broken to U(1)$^2$. We elucidate the geometry of the scalar sector, and also present the superpotentials in both ${\cal N}=2$ and ${\cal N}=1$ supergravity formulation.
\end{abstract}

\pacs{11.25.Tq, 11.15.Yc, 04.65.+e} 

\keywords{AdS/CFT, Chern-Simons, Supergravity, Superpotential}

\maketitle 

\section{Introduction}
\pagenumbering{arabic}




AdS solutions in supergravity theories have played a central role in the framework of AdS/CFT correspondence \cite{Maldacena:1997re}. They provide gravity backgrounds and it is on their boundary where dual field theories can be formulated. Hence, for the case of lower dimensional gauged supergravity theories, AdS solutions which can be uplifted to 10-/11-dimensions are of particular importance. 

Gauged supergravity theories can be obtained through gauging of Poincare supergravity. The gauging procedure  promotes some of the vector fields to non-abelian gauge fields and introduces a gauge coupling constant. As a result, it induces a non-trivial scalar potential, whose critical points may give rise to AdS solutions. 
Gauged supergravity can be also obtained by a consistent truncation of higher dimensional theory.  For consistently truncated models, a solution of lower dimensional theory automatically satisfies the equations of motion of higher dimensional theory. It allows us to embed AdS solutions into string theory.

One of the most well-known examples of consistently truncated gauged supergravity is $D=4$, $\mathcal{N}=8$ SO(8)-gauged theory. Its AdS vacuum solution is uplifted to AdS$_4 \times$ S$^7$ background in M-theory, which is dual to three dimensional superconformal Chern-Simons-matter theory known as ABJM (Aharony-Bergman-Jafferis-Maldacena) theory \cite{Aharony:2008ug}.

There exists a one parameter family of four-dimensional maximal SO(8)$_c$ gauged supergravity \cite{DallAgata:2011aa, DallAgata:2012mfj, DallAgata:2014tph}. Here $c$ is a deformation parameter, i.e. the ratio of a newly introduced magnetic gauge coupling constant $m$ to a usual electric coupling $g$. It is thus also called dyonically gauged theory. With these two coupling constants, the scalar potential of gauged supergravity should exhibit a richer vacuum structure. Hence, it is expected that dyonically gauged supergravity theories might open up new possibilities for AdS vacuum solutions. 

Dyonic gauged supergravity and its application to AdS/CFT correspondence have been studied extensively in a series of papers \cite{Guarino:2015jca, Guarino:2015qaa, Guarino:2015vca}, where 
the authors studied $D=4$, $\mathcal{N}=8$ dyonic ISO(7) gauged supergravity and showed that the four-dimensional theory can be obtained by a consistent truncation of massive type IIA supergravity on S$^6$. The magnetic gauge coupling constant $m$ is identified with the Romans mass of massive type IIA supergravity.
Instead of studying the full $D=4$, $\mathcal{N}=8$ dyonic ISO(7) gauged supergravity theory, one may focus on a suitable subsector of the theory for technical convenience. In the SU(3)-invariant sector \footnote{In the SU(3)-invariant sector, domain wall solutions \cite{Guarino:2016ynd}, BPS black hole solutions \cite{Guarino:2017eag} and graviton mass spectrum \cite{Pang:2017omp} were also studied.}, a new example of AdS/CFT correspondence was proposed by \cite{Guarino:2015jca}. 
 The authors found an $\mathcal{N}=2$ AdS critical point of $D=4$ theory and uplifted to a new $\mathcal{N}=2$ AdS$_4 \times $ S$^6$ solution, which is dual to three-dimensional superconformal Chern-Simons-matter theory
\footnote{This duality was studied further in \cite{Araujo:2016jlx, Araujo:2017hvi}.}. The matter content of the dual field theory consists of a single gauge multiplet with gauge group SU($N$) with a non-zero Chern-Simons level $k$, and three adjoint chiral multiplets enjoying SU(3)-flavor symmetry. They computed the field theoretic partition function using the localization technique and found agreement with the result of the gravitational calculation.
 It was argued in \cite{Fluder:2015eoa} that this correspondence holds for a large class of duality pairs which are inherited from their parent superconformal field theory in four-dimensions, with the same matter contents and quiver diagram. 
   
It is desirable to generalize  this duality to a non-conformal setting. The free energy of the field theory mentioned above is determined by maximizing
\be\label{FE}
F =\dfrac{3^{13/6} \pi}{40} \left(1- \dfrac{i}{\sqrt{3}} \right)\, k^{1/3}\, N^{5/3}
\left(4 \Delta_1 \Delta_2 \Delta_3 \right)^{2/3},
\ee
with respect to the R-charges of the chiral multiplets $\Delta_i$ while keeping the marginality condition of superpotential, i.e. $\Delta_1+\Delta_2+\Delta_3=2$. This procedure is known as $F$-maximization \cite{Jafferis:2010un, Jafferis:2011zi, Pufu:2016zxm}. The explicit form of the free energy with general quiver diagrams and R-charges is given in \cite{Fluder:2015eoa}. Following the ideas of \cite{Freedman:2013ryh, Bobev:2013cja, Bobev:2016nua}, one can think of a general R-charge assignment as a deformation of superconformal theory. This R-charge deformation introduces relevant operators such as mass terms, which induce RG-flows from the superconformal fixed point. Our ultimate goal is to study these RG-flows holographically and reproduce the free energy \eqref{FE} from the supergravity calculation. Since a generic R-charge assignment breaks SU(3) symmetry to U(1)$^2$, our starting point to study the supergravity duals is constructing U(1)$^2$ invariant sector of $\mathcal{N}=8$ dyonic ISO(7) gauged supergravity.

Recently, the U(1)$^2$-invariant sector was constructed using a canonical $\mathcal{N}=2$ formulation in studying BPS black holes \cite{Guarino:2017pkw}. The truncated theory is an $\mathcal{N}=2$ supergravity coupled to three vector multiplets and a hypermultiplet, which is an analogue of STU-model in eleven-dimensional supergravity theory. It is called the dyonic STU model in \cite{Hosseini:2017fjo} and studied further in \cite{Azzurli:2017kxo, Hosseini:2017fjo, Benini:2017oxt} where the Bekenstein-Hawking entropy of the black holes and the topologically twisted index of the dual field theories were computed to give an exact agreement. 



In this note, we revisit the U(1)$^2$-invariant sector of $\mathcal{N}=8$ dyonic ISO(7) gauged supergravity as a preliminary study on the gravity duals of RG-flows induced by a R-charge deformation of Guarino, Jafferis and Valera's (GJV) field theory \cite{kks}. We study the consistent truncation of maximal supergravity to U(1)$^2$-invariant sector and construct the bosonic Lagrangian using the embedding tensor formalism. This U(1)$^2$ truncated theory is also discussed in terms of superpotential and $\mathcal{N}=2$ formulations, respectively. For the neutral scalar sector, we rewrite the bosonic action in an $\mathcal{N}=1$ language by identifying the holomorphic superpotential and the K\"{a}hler potential. It allows us to write down the supersymmetry variations of the fermionic fields, which will be useful in obtaining a set of BPS equations. We also give a brief comment on the critical points of the theory and the scalar mass spectrum around an $\mathcal{N}=2$ fixed point. The appearance of the tachyonic scalar fields indicates that RG-flows can start from this UV fixed point.

 The plan of this paper is as follows. Section \ref{review} briefly reviews the SU(3)-invariant sector of $D=4$ $\mathcal{N}=8$ dyonic ISO(7) gauged supergravity. In section \ref{truncation}, we discuss various aspects of the U(1)$^2$-invariant sector obtained by a consistent truncation of the maximal supergravity in detail. Some useful formulas needed in dyonic gauging
are presented in appendix \ref{append}.

{\bf Note:} As this manuscript was being finalized, we received a paper by A. Guarino \cite{Guarino:2017jly} where the truncation of dyonic gauged supergravity to $\mathcal{N}=2$ supergravity with U(1)$^2$ is also explained in detail. In particular, eqs. (12), (14) of this paper is the same as eqs. (65)-(67) of \cite{Guarino:2017jly}.

\section{Reviews on gauged supergravity}\label{review}
\subsection{$D=4$ dyonic ISO(7) gauged maximal supergravity}
In this section, we briefly review $D=4$ dyonic ISO(7) gauged maximal supergravity. 
The bosonic field contents are metric, scalars ${\mathcal{V}}_{\mathbb M}^{\phantom M \mathbb{\underline{N}}}$ and vectors ${\mathcal{A}^\mathbb{M}}$
\footnote{In addition, there is an auxiliary two-form tensor field ${\mathcal B}_\alpha$ in a dyonic gauged theory. For the detailed reviews, one can refer to \cite{Guarino:2015qaa}.}.

The 70 scalar fields of $D=4$ maximal supergravity parametrize the coset E$_{7(7)}/$SU$(8)$.
First, let us consider E$_{7(7)}$ generators $[{t_\alpha}]_\mathbb{M}^{\phantom {\mathbb{M}} \mathbb{N}}$.
Here the index $\mathbb{M}=1, 2, \cdots, 56$ is a fundamental index of E$_{7(7)}$ and
$\alpha =1, 2, \cdots, 133$  is an adjoint index of E$_{7(7)}$.
The adjoint representation $\mathbf{133}$ of E$_{7(7)}$ is decomposed under SL(8) to
$\mathbf{133} \rightarrow \mathbf{63}+\mathbf{70}$. Hence,
the generators of E$_{7(7)}$ are
$t_\alpha=t_{M}^{\phantom A N} \oplus t_{MNPQ}$ with $t_\alpha=t_{M}^{\phantom A M}=0$ and $t_{[MNPQ]}$
in SL(8) basis. Here $M =1, \cdots, 8$ is a fundamental index of SL(8).
The adjoint representation $\mathbf{63}$ of SL(8) is decomposed under SO(8) to anti-symmetric $[A,B]$ and symmetric-traceless $(A,B)$ representations. The representation $\mathbf{70}$ of SL(8) is decomposed under SO(8) to totally anti-symmetric anti-self-dual $[A,B,C,D]_-$ and self-dual $[A,B,C,D]_+$ representations.
\be
\mathbf{133}
\quad \underset{SL(8)}{\longrightarrow} \quad  \mathbf{63}+\mathbf{70}
\quad \underset{SO(8)}{\longrightarrow} \quad \left(\mathbf{28}+\mathbf{35}_v \right) +\left(\mathbf{35}_s+\mathbf{35}_c \right).
\ee
The adjoint representation $\mathbf{63}$ of SU(8) is decomposed under SO(8) to anti-symmetric $[I,J]$ and symmetric-traceless $(I,J)$ representations.
\be
\mathbf{63}
\quad \underset{SO(8)}{\longrightarrow} \quad  \mathbf{28}+\mathbf{35}_s.
\ee
As a result, the 70 scalar fields which parametrize the coset E$_{7(7)}/$SU$(8)$ are 
$\mathbf{35}_c+\mathbf{35}_v$. They are represented by
totally anti-symmetric $[A,B,C,D]_{+}$ and symmetric-traceless $(A,B)$, respectively. The three inequivalent $\mathbf{35}$ representations of SO(8) are summarized below
\footnote{We follow the notation of \cite{Duff:1986hr}. (See page 67.)}.
\begin{displaymath}
\begin{array}{c|ccc}
\hline
 & s & c & v \\
\hline\hline
\mathbf{35}_s  & (IJ)     & [I'J'K'L']_+ & [ABCD]_- \\
\mathbf{35}_c  & [IJKL]_- & (I'J')       & [ABCD]_+ \\
\mathbf{35}_v  & [IJKL]_+ & [I'J'K'L']_- & (AB)\\
\hline
\end{array}
\end{displaymath}
Here $I, J= 1, \cdots, 8$ are for SO(8) spinor indices, and $I', J'= 1, \cdots, 8$ for conjugate indices, $A, B= 1, \cdots, 8$ for vector indices. $\pm$ imply the (anti-)self-duality condition. 

The fundamental representation $\mathbf{56}$ of E$_{7(7)}$ is decomposed under SL(8) as $\mathbf{56} \rightarrow \mathbf{28}+\mathbf{28'}$. This implies that vector can be separated into the electric and magnetic components as
${\mathcal{A}^\mathbb{M}} \rightarrow \mathcal{A}^{[MN]} \oplus \mathcal{A}_{[MN]}$.

Now we consider gauged supergravity using the embedding tensor formalism. Gauging procedure implies that we choose some of vector fields and promote them to non-Abelian gauge fields. All the information of gauging can be encoded in the embedding tensor $\Theta_{\mathbb{M}}^{\phantom M \alpha}$. It specifies the vector fields and the generators which participate in gauging. The electric and magnetic parts of the embedding tensor have forms of
\be
\Theta_{[MN] \phantom{P}Q}^{\phantom{[MN]}P}=2 \delta_{[M}^P \theta_{N]Q}
\quad
\textrm{and}
\quad
\Theta^{[MN]P}_{\phantom{[MN]P}Q}=2 \delta_{Q}^{[M}\xi^{N]P}.
\ee
Here the E$_{7(7)}$ adjoint index $\alpha$ is restricted to the adjoint of SL(8) since we consider gauge group G $\subset$ SL(8) $\subset$ E$_{7(7)}$.
For the case of dyonic ISO(7) gauging, we have
\be
\theta=
\left(\begin{array}{cc} 
 \mathbb{I}_{7} & 0 \\
 0 & 0
 \end{array} \right)
\quad 
\textrm{and}
\quad
\xi=
\left(\begin{array}{cc} 
 0_{7 \times 7} & 0 \\
 0 & c
 \end{array} \right),
\ee
where $c \equiv m/g$
 is the ratio of a magnetic gauge coupling $m$ to a electric gauge coupling $g$.
\subsection{The SU(3) invariant sector}
Now we move on to review the SU(3) invariant sector of $D=4$ $\mathcal{N}=8$ dyonic ISO(7) gauged supergravity. We consistently truncate the $\mathcal{N}=8$ theory to $\mathcal{N}=2$ sub-sector, which contains SU(3) singlets only.
First, the embedding SU(3) into SO(7) $\subset$ ISO(7)$\equiv$ SO(7)  $ \ltimes \mathbb{R}^7$ is as follows 
(See eq. 3.2 in \cite{Guarino:2015qaa}
\footnote{The branching rules can be found in \cite{Bobev:2010ib}. (See eq. C.2-C.7.)}.)
\begin{equation}
\mathrm {SO(7)} \supset \mathrm {SO(6)} \sim \mathrm {SU(4)} \supset \mathrm {SU(3)}.
\end{equation}
Let us begin by the decomposion of $\mathbf{8}_v$ of SO(8)
\footnote{Note that the SU(4) embedding is different from that of Freedman-Pufu
\cite{Freedman:2013ryh}, where $\mathbf{8}_s$ is decomposed under SU(4) as
\be
\mathbf{8}_s
\quad \underset{SU(4)}{\longrightarrow} \quad \mathbf{6}+\mathbf{1}+\mathbf{1}. \nn
\ee 
Then the supergravity scalars $\mathbf{35}_c,\, \mathbf{35}_v$ are represented by $[I,J,K,L]_- ,\, [I,J,K,L]_+$. They are decomposed in a same way as
\be
\mathbf{35}_{c, v}
\quad \underset{SU(4)}{\longrightarrow}\quad \mathbf{15}+\mathbf{10}+\mathbf{\bar{10}}. \nn
\ee 
}
as
\be
\mathbf{8}_v 
\quad \underset{SO(7)}{\longrightarrow} \quad  \mathbf{7}+\mathbf{1}
\quad \underset{SU(4)}{\longrightarrow} \quad \mathbf{6}+\mathbf{1}+\mathbf{1}
\quad \underset{SU(3)}{\longrightarrow} \quad \mathbf{\bar{3}}+\mathbf{3}+\mathbf{1}+\mathbf{1}.
\ee
Then, the 70 scalars $\mathbf{35}_c$ and $\mathbf{35}_v$, i.e. $[A,B,C,D]_{+}$ and $(A,B)$, respectively, can be decomposed under $\mathrm {SO(6)} \sim \mathrm {SU(4)}$
as
\begin{align}
\mathbf{35}_{c}
\quad &\underset{SU(4)}{\longrightarrow} \quad \mathbf{15}+\mathbf{10}+\mathbf{\bar{10}}. \nn\\
\mathbf{35}_{v}
\quad &\underset{SU(4)}{\longrightarrow} \quad \mathbf{20'}+\mathbf{6}+\mathbf{6}+
\mathbf{1}+\mathbf{1}+\mathbf{1}. \nn
\end{align}
The decomposition of scalar and vector fields under $\mathrm{SU(3)} \subset \mathrm{SU(4)}$ are summarized in the table. Here $\phi_{(A,B)},\,\phi_{[A,B,C,D]_+}$ are $\mathbf{35}_{v},\,\mathbf{35}_{c}$ scalars, respectively.
\begin{displaymath}
\begin{array}{cc|cc|c}
\hline
&SO(8)& &SO(6) & SU(3) \\
\hline
\phi_{(A,B)}&\mathbf{35}_{v}& \phi_{ab}&\mathbf{20'} & \mathbf{8}+\mathbf{\bar{6}}+\mathbf{6}\\
&& \phi_{a1}&\mathbf{6} & \mathbf{\bar{3}}+\mathbf{3} \\
&& \phi_{a8}&\mathbf{6} & \mathbf{\bar{3}}+\mathbf{3}\\
&& \phi_{11}&\mathbf{1} & \mathbf{1}\\
&& \phi_{18}&\mathbf{1} & \mathbf{1}\\
&& \phi_{88}&\mathbf{1} & \mathbf{1}\\
\hline\hline
\phi_{[A,B,C,D]_+}&\mathbf{35}_{c}&\phi_{ab18}& \mathbf{15} & \mathbf{8}+\mathbf{3}+\mathbf{\bar{3}}+\mathbf{1} \\
&&\phi_{abc1}& \mathbf{20} & \mathbf{1}+\mathbf{1}+\mathbf{6}+\mathbf{\bar{6}}+\mathbf{3}+\mathbf{\bar{3}}\\
\hline\hline
\mathcal{A}^{AB}&\mathbf{28}& A^{ab} & \mathbf{15} & \mathbf{8}+\mathbf{3}+\mathbf{\bar{3}}+\mathbf{1}\\
&& A^{a1} & \mathbf{6} & \mathbf{3}+\mathbf{\bar{3}}\\
&& A^{a8} & \mathbf{6} & \mathbf{3}+\mathbf{\bar{3}}\\
&& A^{18} & \mathbf{1} & \mathbf{1}\end{array}
\end{displaymath}
Here, $a=2, \cdots, 7$ is a fundamental index of SO(6).
As a result, there are six real scalar fields, one gauge field and one graviphoton in the SU(3)-invariant sector. The truncated theory is an ${\mathcal N}=2$ supergravity coupled to one vector multiplet and one hypermultiplet.

At the ${\mathcal N}=2$ U(3) fixed point, the spectra of scalar and vector fields are summarized as follows. (See Table 3 of \cite{Guarino:2015qaa}.)
\begin{displaymath}
\begin{array}{c||c}
& M^2 L^2\\
\hline
\textrm{scalar}& 3 \pm \sqrt{17},\,2,\,2,\,2,\,0 \\
\textrm{vector}& 4,\,0
\end{array}
\end{displaymath}
They can be organized into an
${\mathcal N}=2$ long vector multiplet. See, for example, the table 1 of \cite{Nicolai:1985hs}, the table 2 of \cite{Ahn:2008ya}, the table 15 of \cite{Klebanov:2008vq}, the table A.6 of \cite{Bobev:2010ib} and the table 4 of \cite{Pang:2017omp}. One can easily compute the masses of vector and scalars of ${\mathcal N}=2$ long vector multiplet using the AdS/CFT dictionary $M^2 L^2 =\left(\Delta-p \right)\left(\Delta+p-3 \right)$
with
$E_0=\frac{1}{2}\left(1+\sqrt{17} \right)$.
\begin{displaymath}
\begin{array}{cc||c||ccccc}
\hline
\textrm{spin}& p & 1 & && 0 \\
\hline
\textrm{energy}& \Delta & E_0+1 & E_0+2 & E_0+1 & E_0+1 & E_0+1 & E_0\\
\hline\hline
\textrm{mass} & M^2 L^2 & 4 & 3+\sqrt{17} & 2 & 2 & 2 & 3-\sqrt{17}\\
\hline
\end{array}
\end{displaymath}
The massless vector is a graviphoton and the massless scalar is a Goldstone boson to be eaten. 
\section{Consistent truncation}\label{truncation}
\subsection{The $\mathrm U(1)^2$-invariant sector}
We study the $\mathrm U(1)^2$ truncation of $D=4$, $\mathcal{N}=8$ dyonic ISO(7) gauged supergravity. We are looking for the scalar fields invariant under $\mathrm U(1)^2$ i.e. the two Cartan generators of $\mathrm {SU(3)}$ as
\begin{displaymath}
\lambda_3=\left(\begin{array}{cccccc}
0 & -1 & 0 & 0 & 0 & 0 \\
1 &  0 & 0 & 0 & 0 & 0 \\
0 &  0 & 0 & 1 & 0 & 0 \\
0 &  0 &-1 & 0 & 0 & 0 \\
0 &  0 & 0 & 0 & 0 & 0 \\
0 &  0 & 0 & 0 & 0 & 0 
\end{array}\right), \qquad
\lambda_8=\left(\begin{array}{cccccc}
0 & -1 & 0 & 0 & 0 & 0 \\
1 &  0 & 0 & 0 & 0 & 0 \\
0 &  0 & 0 &-1 & 0 & 0 \\
0 &  0 & 1 & 0 & 0 & 0 \\
0 &  0 & 0 & 0 & 0 & 2 \\
0 &  0 & 0 & 0 &-2 & 0 
\end{array}\right).
\end{displaymath}

First, we obtain five real scalar fields, which are invariant under the action of $\lambda_3$ and $\lambda_8$, in $\mathbf{35}_c$ as
\be
\phi_{2318}, \quad \phi_{4518}, \quad \phi_{6718}, \quad \phi_{2461}, \quad \phi_{2471}.
\ee
Let us consider the following expansions
\begin{align}
\dfrac{1}{2}\, &\phi_{ab18}\, e^a \wedge e^b
=\alpha\, J + \dfrac{1}{2}\, f^m\, \lambda^m_{ab}\, e^a \wedge e^b+ \cdots,\nn\\
\dfrac{1}{3!}\, &\phi_{abc1}\, e^a \wedge e^b \wedge e^c
=\beta\, \Omega +\gamma\, \bar{\Omega}+ \cdots,
\end{align}
where
\begin{align}
J &= e^2\wedge e^3 + e^4 \wedge e^5 + e^6 \wedge e^7, \nn\\
\Omega &= \left(e^2 + i\, e^3 \right) \wedge \left(e^4 + i\, e^5 \right) \wedge \left(e^6 + i\, e^7 \right).
\end{align}
Hence, we conclude that $\alpha,\, \beta,\, \gamma$ are $\mathrm{SU(3)}$ singlets and $f^3, f^8$ are new singlets in the $\mathrm U(1)^2$-invariant sector.
\begin{align}
\alpha &= \dfrac{1}{3} \left(\phi_{2318}+\phi_{4518}+\phi_{6718} \right),\nn\\
f^3 &=-\dfrac{1}{2}\left(\phi_{2318}-\phi_{4518} \right),\nn\\
f^8 &=-\dfrac{1}{6} \left(\phi_{2318}+\phi_{4518}-2\,\phi_{6718} \right),\nn\\
\beta &= \dfrac{1}{2}\left(\phi_{2461}- i \,\phi_{2471} \right),\nn\\
\gamma &= \dfrac{1}{2}\left(\phi_{2461}+ i \,\phi_{2471} \right).
\end{align}
With the self-duality conditions, we have the following relations between various scalars.
\begin{align}
\phi_{2318}&=\phi_{4567}, \qquad \phi_{4518}=\phi_{2367}, \qquad \phi_{6718}=\phi_{2345}, \nn\\
\phi_{2461}&=-\phi_{2571}=-\phi_{3471}=-\phi_{3561}
           = \phi_{3578}=-\phi_{3468}=-\phi_{2568}=-\phi_{2478}, \nn\\
\phi_{2471}&=\phi_{2561}=\phi_{3461}=-\phi_{3571}=-\phi_{3568}
=-\phi_{3478}=-\phi_{2578}=\phi_{2468}. \nn
\end{align}

In $\mathbf{35}_v$, there are three SU(3) singlets $\phi_{11},\,\phi_{18},\,\phi_{88}$. Imposing U(1)$^2$-invariance gives $\phi_{22}=\phi_{33},\, \phi_{44}=\phi_{55},\, \phi_{66}=\phi_{77}$. A traceless condition of $\phi_{ab}$ gives $\phi_{66}=-\phi_{22}-\phi_{44}.$
As a result of further U(1)$^2$ truncation, we obtain two more singlets 
$\phi_{22},\,\phi_{44}
$.

Now we construct the bosonic Lagrangian explicitly\footnote{We closely follow the procedure of the SU(3) truncation of \cite{Guarino:2015qaa}. See Sec. 3.1. }. First, we provide the generators of E$_{7(7)}$ which are invariant under the U(1)$^2$. They are
\begin{align}
g_1 &= -t_{2}^{\phantom 2 2} - t_{3}^{\phantom 2 3} + t_{4}^{\phantom 2 4} + t_{5}^{\phantom 2 5} + t_{6}^{\phantom 2 6} + t_{7}^{\phantom 2 7}-(t_{1}^{\phantom 2 1}+t_{8}^{\phantom 2 8}),\nn\\
g_2 &=t_{2}^{\phantom 2 2} + t_{3}^{\phantom 2 3} - t_{4}^{\phantom 2 4} - t_{5}^{\phantom 2 5} + t_{6}^{\phantom 2 6} + t_{7}^{\phantom 2 7}-(t_{1}^{\phantom 2 1}+t_{8}^{\phantom 2 8}),\nn\\
g_3 &=t_{2}^{\phantom 2 2} + t_{3}^{\phantom 2 3} + t_{4}^{\phantom 2 4} + t_{5}^{\phantom 2 5} - t_{6}^{\phantom 2 6} - t_{7}^{\phantom 2 7}-(t_{1}^{\phantom 2 1}+t_{8}^{\phantom 2 8}),\nn\\
g_4 &= g_4^{-}+g_4^{+}=  (t_{1}^{\phantom 2 8})+ (t_{8}^{\phantom 2 1}),\nn\\
g_5 &=  t_{1}^{\phantom 2 1}- t_{8}^{\phantom 2 8},\nn\\
g_6 &= g_6^{-}+g_6^{+}=(t_{4567})+(t_{1238}),\nn\\
g_7 &= g_7^{-}+g_7^{+}=(t_{2367})+(t_{1458}),\nn\\
g_8 &= g_8^{-}+g_8^{+}=(t_{2345})+(t_{1678}),\nn\\
g_9 &= g_9^{-}+g_9^{+}=(t_{1246}-t_{1257}-t_{1347}-t_{1356})+(t_{8357}-t_{8346}-t_{8256}-t_{8247}),\nn\\
g_{10} &= g_{10}^{-}+g_{10}^{+}=(t_{3571}-t_{3461}-t_{2561}-t_{2471})+(t_{8246}-t_{8257}-t_{8347}-t_{8356}).
\end{align}
The explicit form of the generators $t_A^{\phantom A B}$ as $56 \times 56$ matrices is recorded in the appendix \eqref{gen}. The generators $g_4,\, g_5,\, g_9,\, g_{10},\, g_1+g_2+g_3,\, g_6+g_7+g_8$ are the same as 
those of the SU(3) invariant truncation (eq. 3.5 in \cite{Guarino:2015qaa}). Here we add four generators: $g_2,\, g_3,\, g_7,\, g_8$. 
By exponetiating the generators, we construct the coset representative in the SL(8) basis as
\begin{align}
{\cal{V}}&= 
\textrm{exp}\left( a\, g_4^+ -6\,\zeta\, g_9^+ -6\,\tilde{\zeta}\, g_{10}^+
-12\, \chi_1\, g_6^+ -12\, \chi_2\, g_7^+ 
-12\, \chi_3\, g_8^+
 \right)\nn\\
&\times \textrm{exp} \left( \dfrac{\varphi_1}{4} \,g_1 
+\dfrac{\varphi_2}{4}\, g_2 +\dfrac{\varphi_3}{4}\,g_3
+\phi \,g_5\right).
\end{align}
Here, $g_1,\, g_2,\, g_3,\, g_5$ are Cartan generators and 
$g_4^{+},\, g_6^{+},\, g_7^{+},\, g_8^{+},\, g_9^{+},\, g_{10}^{+}$ are the positive root generators. 
$(\varphi_n,\, \phi)$ are dilatonic scalars and $(a,\, \zeta,\, \tilde{\zeta},\chi_n)$ are axionic scalars ($n=1,2,3$).
This coset representative can be factorized to ${\cal{V}}={\cal{V}}_{\textrm{SK}}{\cal{V}}_{\textrm{QK}}$
\footnote{Here, we used $g_4^+,\, g_9^+,\, g_{10}^+$ commute with $g_1,\, g_2,\, g_3,\, g_6^+,\, g_7^+,\, g_8^+$ and 
$[g_1,g_7^+]=[g_1,g_8^+]=[g_2,g_6^+]=[g_2,g_8^+]=[g_3,g_6^+]=[g_3,g_7^+]=0.$} 

\begin{align}
{\cal{V}}_\textrm{SK}&=\left(e^{-12\, \chi_1\, g_6^+} e^{\frac{\varphi_1}{4} \,g_1}\right)
\left(e^{-12\, \chi_2\, g_7^+} e^{\frac{\varphi_2}{4} \,g_2}\right)
\left(e^{-12\, \chi_3\, g_8^+} e^{\frac{\varphi_3}{4} \,g_3}\right),\nn\\
{\cal{V}}_\textrm{QK}&=e^{a\, g_4^+ -6\,\zeta\, g_9^+ -6\,\tilde{\zeta}\, g_{10}^+}\,e^{\phi \,g_5}
.
\end{align}
The six real scalars $(\chi_n,\, \varphi_n)$ parametrize a special K\"{a}hler manifold ${\cal{M}_\textrm{SK}}=[\textrm{SU}(1,1)/\textrm{U}(1)]^3$ and the four real scalars $(a,\, \zeta,\,\tilde{\zeta},\, \phi)$ parametrize a quaternionic K\"{a}hler manifold
 ${\cal{M}_\textrm{QK}}=\textrm{SU}(2,1)/(\textrm{SU}(2) \times \textrm{U}(1))$.
Using the scalar matrix ${\cal{M}}$ defined as ${\cal{M}}={\cal{V}} {\cal{V}}^t$, we can write down the scalar kinetic terms
\footnote{
The covariant derivative of the scalar matrix ${\mathcal{M}}$ is 
\begin{align}
D_\mu {\mathcal{M}}_{\mathbb {MN}} &=\partial_\mu {\mathcal{M}}_{\mathbb {MN}} 
- g A_\mu ^{\phantom{\mu}\mathbb {P}} X_{\mathbb {PM}}^{\phantom{PM}\mathbb {Q}} 
{\mathcal{M}}_{\mathbb {QN}} 
- g A_\mu ^{\phantom{\mu}\mathbb {P}} X_{\mathbb {PN}}^{\phantom{PN}\mathbb {Q}} 
{\mathcal{M}}_{\mathbb {MQ}},\nn\\
D_\mu {\mathcal{M}}^{-1} &= D_\mu {\mathcal{M}}^{\mathbb {MN}} = \Omega^{\mathbb {MP}}\Omega^{\mathbb {NQ}}D_\mu {\mathcal{M}}_{\mathbb {PQ}},\nn
\end{align}
where $\Omega^{\mathbb {MN}}=
\left(\begin{array}{cc} 
 0_{28} & \mathbb{I}_{28} \\
 -\mathbb{I}_{28} & 0_{28}
 \end{array} \right).$
 See eq. 4.51 of \cite{deWit:2007kvg}.
} 
\begin{align}\label{Lscalar}
e^{-1}{\cal{L}}_{\textrm{scalar}}^{\textrm{kin}} \equiv
& \dfrac{1}{48}\, \textrm{Tr}\, D_\mu {\mathcal{M}}D^\mu {\mathcal{M}}^{-1},\nn \\
=&
-\dfrac{1}{2} \sum_{n=1}^3 \left((\partial_\mu \varphi_n)^2 + e^{2\varphi_n}(\partial_\mu \chi_n)^2\right)
-2 (\partial_\mu \phi)^2 \\
&-\dfrac{1}{2} e^{4\phi} \left( D_\mu a +\dfrac{1}{2}
 \left(\zeta D_\mu \tilde{\zeta}- \tilde{\zeta} D_\mu \zeta  \right)\right)^2
 -\dfrac{1}{2} e^{2\phi} \left((D_\mu \zeta)^2 +(D_\mu \tilde{\zeta})^2 \right) , \nn
\end{align}
where the covariant derivatives are 
\begin{align}\label{cd}
D_\mu \zeta &= \partial_\mu \zeta -g\,(A^1+A^2+A^3)\, \tilde{\zeta},\nn\\
D_\mu \tilde{\zeta} &= \partial_\mu \tilde{\zeta} +g\,(A^1+A^2+A^3)\, \zeta, \nn\\
D_\mu a &= \partial_\mu a + g A^0 -m \tilde{A}_0.
\end{align}
Here we used the so-called X-tensors defined as
\be
X_{\mathbb{MN}}^{\phantom{\mathbb{MN}}\mathbb{P}}
\equiv\Theta_{\mathbb{M}}^{\phantom M \alpha} [t_\alpha]_{\mathbb{N}}^{\phantom{\mathbb{N}}\mathbb{P}}.
\ee
They contain all the information about gauging process through the embedding tensors. The explicit form of X-tensors are given by \eqref{xtensorform}. With the scalar matrix ${\mathcal{M}}$ and X-tensors, one can calculate the scalar potential.
The scalar potential has a rather long expression, and can be written as follows.
\begin{align}\label{pot}
V& \equiv \dfrac{g^2}{168} \left(X_{\mathbb {MN}}^{\phantom{MN}\mathbb R}X_{\mathbb{PQ}}^{\phantom{MN}\mathbb S} 
{\cal{M}}^{\mathbb{MP}} {\cal{M}}^{\mathbb{NQ}} {\cal{M}}_{\mathbb{RS}}
 +7 X_{\mathbb{MN}}^{\phantom{MN}\mathbb Q}X_{\mathbb{PQ}}^{\phantom{MN}\mathbb N} 
{\cal{M}}^{\mathbb{MP}}  \right),\nn\\
&=g^2 \Big[ \dfrac{1}{8} (\zeta^2+\tilde{\zeta}^2)^2 \, e^{-\varphi_1-\varphi_2-\varphi_3+4\phi} \nn\\
&\quad\quad\quad\quad\times \left( (e^{2\varphi_1+2\varphi_2}+e^{2\varphi_2+2\varphi_3}+e^{2\varphi_3+2\varphi_1})+e^{2\varphi_1+2\varphi_2+2\varphi_3}(\chi_1+\chi_2+\chi_3)^2\right) \nn\\
&\phantom{g^2 \Big[} +\dfrac{1}{2} (\zeta^2+\tilde{\zeta}^2) \, e^{-\varphi_1-\varphi_2-\varphi_3+4\phi} \nn\\
&\quad\quad\quad\quad\times 
\Big\{
e^{-2\phi}\left( (e^{2\varphi_1+2\varphi_2}+e^{2\varphi_2+2\varphi_3}+e^{2\varphi_3+2\varphi_1})
-2e^{\varphi_1+\varphi_2+\varphi_3}(e^{\varphi_1}+e^{\varphi_2}+e^{\varphi_3})\right) \nn\\
&\quad\quad\quad\quad\quad
+e^{2\varphi_1+2\varphi_2+2\varphi_3-2\phi}(\chi_1+\chi_2+\chi_3)^2 \nn\\
&\quad\quad\quad\quad\quad
+\left(e^{2\varphi_1+2\varphi_2}\chi_1\chi_2+e^{2\varphi_2+2\varphi_3}\chi_2\chi_3+e^{2\varphi_3+2\varphi_1}\chi_3\chi_1\right)\nn\\
&\quad\quad\quad\quad\quad
+e^{2\varphi_1+2\varphi_2+2\varphi_3}\chi_1\chi_2\chi_3(\chi_1+\chi_2+\chi_3) \Big\}\nn\\
&\phantom{g^2 \Big[}
-4(e^{\varphi_1}+e^{\varphi_2}+e^{\varphi_3})
-2 e^{2\phi} (e^{-\varphi_1}+e^{-\varphi_2}+e^{-\varphi_3}+e^{\varphi_1}\chi_1^2+e^{\varphi_2}\chi_2^2+e^{\varphi_3}\chi_3^2)\nn\\
&\phantom{g^2 \Big[}
+\dfrac{1}{2}e^{-\varphi_1-\varphi_2-\varphi_3+4\phi}
(1+e^{2\varphi_1}\chi_1^2)(1+e^{2\varphi_2}\chi_2^2)(1+e^{2\varphi_3}\chi_3^2) \Big]\nn\\
&- \dfrac{g\, m}{2} e^{\varphi_1+\varphi_2+\varphi_3+4\phi} \left( (\zeta^2+\tilde{\zeta}^2)(\chi_1+\chi_2+\chi_3)+2 \chi_1 \chi_2 \chi_3 \right)\nn\\
&+ \dfrac{m^2}{2} e^{\varphi_1+\varphi_2+\varphi_3+4\phi}.
\end{align}
The scalar potential depends on two gauge coupling constants $g, \,m$ and eight real scalar fields $(\chi_n, \, \phi_n, \, \phi, \, \rho)
\footnote{With the redefinition of the fields $\tilde{\zeta}+ i\, \zeta= 2\,\rho\,e^{i\beta}$, the scalar potential depend only on $\rho^2=\frac{1}{4} \left(\tilde{\zeta}^2+\zeta^2 \right)$.}
$.
When we have $\varphi \equiv \varphi_1=\varphi_2=\varphi_3$ and $\chi \equiv \chi_1=\chi_2=\chi_3$, then the scalar potential \eqref{pot} reduces to that of the SU(3) invariant theory. See e.g. eq. (3.11) of \cite{Guarino:2015qaa}.
\subsection{Superpotential formulation}
We introduce the complex coordinates $t_n$ and $u$ as
\be
t_n =-\chi_n + i \,e^{-\varphi_n} \qquad \textrm{and} \qquad u= \rho +i\, e^{-\phi},
\ee
which parameterize the upper-half complex planes, respectively. Then, we perform one more coordinate transformation to the Poincar\'{e} unit disk
\be
z_n = \dfrac{t_n - i}{t_n +i} \qquad \textrm{and} \qquad \zeta_{12} =\dfrac{u-i}{u+i}.
\ee
We have found the following superpotential.
\begin{align}
\mathcal{W}
&=\dfrac{1}{2}\,(1-|z_1|^2)^{-\frac{1}{2}} (1-|z_2|^2)^{-\frac{1}{2}} (1-|z_3|^2)^{-\frac{1}{2}} (1-|\zeta_{12}|^2)^{-2}\nn\\
&\phantom{=}\times \Big(  \dfrac{i}{4} (1-\zeta_{12})^4 (1-z_1)(1-z_2)(1-z_3)\Big)\nn\\
&\phantom{=}\times \Big[ m-i\, g \dfrac{(1+z_1)(1+z_2)(1+z_3)}{(1-z_1)(1-z_2)(1-z_3)}
-2i\, g\left(\dfrac{1+z_1}{1-z_1}+\dfrac{1+z_2}{1-z_2}+\dfrac{1+z_3}{1-z_3} \right) \dfrac{(1+\zeta_{12})^2}{(1-\zeta_{12})^2}\Big].
\end{align}
This superpotential is related to the scalar potential \eqref{pot} through the usual formula,
\begin{align}\label{SP}
    V 
    = 8 \left( \sum_{n=1}^34 (1-|z_n|^2)^2 \left|\dfrac{\partial W}{\partial z_n}\right|^2
        +(1-|\zeta_{12}|^{2})^2 \left|\dfrac{\partial W}{\partial \zeta_{12}}\right|^2
        -3 W^2  \right),
\end{align} 
where we have two superpotentials $W =|\mathcal{W}_+|\equiv |\mathcal{W}(z_n,\zeta_{12})|$ or $W =|\mathcal{W}_-|\equiv |\mathcal{W}(z_n,\bar{\zeta}_{12})|$. Note that $|\mathcal{W}_+|^2 \neq |\mathcal{W}_-|^2$.

\subsection{An ${\cal{N}}=2$ formulation}
In this subsection, we rewrite the bosonic action in a canonical ${\cal N}=2$ form \cite{Andrianopoli:1996cm, Trigiante:2016mnt} 
\footnote{The contents of this section overlap with the previous studies in \cite{Guarino:2017pkw, Hosseini:2017fjo, Benini:2017oxt}.}. The ingredients, which determine the supergravity action, are the holomorphic section $X^\Lambda$ and the prepotential ${\mathcal{F}}$ of the special K\"{a}hler manifold of the vector multiplet scalars and the metric $h_{uv}$ and the moment maps $P_\alpha^{\phantom{\alpha}x}$ of the quaternionic manifold of the hypermultiplet scalars.

First let us study the special K\"{a}hler manifold of the vector multiplet scalars.
 The prepotential is
\be
{\mathcal{F}} = -2 \sqrt{X^0 X^1 X^2 X^3},
\ee
with the holomorphic section $X^M=(X^\Lambda, F_\Lambda)$
\be
X^M=(-t_1\, t_2\, t_3,\, -t_1,\, -t_2,\, -t_3,\, 1,\, t_2\, t_3,\, t_1\, t_3,\, t_1\, t_2).
\ee
Note that once we choose $X^\Lambda$, then $F_\Lambda$ are determined by $F_\Lambda = \partial {\mathcal{F}}/ \partial X^\Lambda$.
Here, $M=1, \cdots, 2(n_v+1)$ is an Sp$(2n_v+2,\mathbb{R})$ vector index, $\Lambda=0, \cdots, n_v$ where $n_v$ is the number of the vector multiplets, i.e. $n_v=3$ for our theory and $t_1, t_2, t_3$ are coordinates of $[\textrm{SU}(1,1)/\textrm{U}(1)]^3$ manifold. 
The K\"{a}hler potential is
\begin{align}
K&=- \textrm{log} \left(i \bar{X}^M \Omega_{MN}X^N \right),\nn\\
& =- \textrm{log} \left(i (t_1-\bar{t}_1) (t_2-\bar{t}_2) (t_3-\bar{t}_3)\right).
\end{align}
The metric is written as
\be
-\sum_{n=1}^3 K_{t_n \bar{t}_n}\, dt_n\, d\bar{t}_n = -\dfrac{1}{4}\sum_{i=n}^3 
\left(d\varphi_n^2 +e^{2\varphi_n} d\chi_n^2 \right).
\ee
The components of the vielbein 
$f_{t_n}^{\phantom{t_n}M}$ and $\bar{f}_{\bar{t}_n}^{\phantom{t_n}M}$ are
\footnote{The definition of the vielbein  and its complex conjugate (See, for example, eq. 3.34 in \cite{Guarino:2015qaa}) is 
\be
f_{t_n}^{\phantom{t_n}M} \equiv 
\partial_{t_n} \left(e^{K/2} X^M \right)+\dfrac{1}{2}e^{K/2} X^M \partial_{t_n} K,
\quad
\bar{f}_{\bar{t}_n}^{\phantom{t_n}M} \equiv 
\partial_{\bar{t}_n} \left(e^{K/2} \bar{X}^M \right)+\dfrac{1}{2}e^{K/2} \bar{X}^M \partial_{\bar{t}_n} K.\nn
\ee}
, for example,
\begin{align}
f_{t_1}^{\phantom{t_i}M}=&\dfrac{1}{(t_1-\bar{t}_1)\left(i (t_1-\bar{t}_1)(t_2-\bar{t}_2)(t_3-\bar{t}_3)\right)^{1/2}}\nn\\
&\times \left(\bar{t}_1\, t_2\, t_3,\, \bar{t}_1,\, t_2,\, t_3,\, -1,\, -t_2\,t_3,\, -\bar{t}_1\, t_3,\, -\bar{t}_1\, t_2 \right),\nn\\
\bar{f}_{\bar{t}_1}^{\phantom{t_i}M}=&\dfrac{1}{(t_1-\bar{t}_1)\left(i (t_1-\bar{t}_1)(t_2-\bar{t}_2)(t_3-\bar{t}_3)\right)^{1/2}}\nn\\
&\times \left(-t_1\, \bar{t}_2\, \bar{t}_3,\, -t_1,\, -\bar{t}_2,\, -\bar{t}_3,\, 1,\, \bar{t}_2\,\bar{t}_3,\, t_1\, \bar{t}_3,\, t_1\, \bar{t}_2 \right).
\end{align}
One can easily obtain other components $f_{t_2}^{\phantom{t_i}M},\, f_{t_3}^{\phantom{t_i}M},\,\bar{f}_{\bar{t}_2}^{\phantom{t_i}M}$ and $\bar{f}_{\bar{t}_3}^{\phantom{t_i}M}$.

Now let us consider the quaternionic manifold of the hypermultiplet scalars.
Here $u=1,\cdots, 4n_h$ where $n_h$ is the number of the hypermultiplets i.e. $n_h=1$, $\alpha=1, \cdots, 8$ is the adjoint index of SU$(2,1)$ and $x=1,2,3$.
From the covariant derivatives of the hypermultiplet scalars \eqref{cd}, we obtain
\be
D_{\mu}= \partial_{\mu}+g\left( A^0 - c\tilde{A}_0\right)\partial_a
+ g \left(A^1+A^2+A^3 \right)
\left(\zeta \partial_{\tilde{\zeta}}-\tilde{\zeta }\partial_{\zeta} \right).
\ee
Then, one can easily read off the Killing vectors\footnote{
The covariant derivative of the scalars in the hypermultiplets is
\be
D_\mu q^u = \partial_\mu q^u +A_\mu^{\phantom{\mu}M} \Theta_M^{\phantom{M}\alpha} k_{\alpha}^{\phantom{\alpha}\nu} \partial_{\nu} q^u,\nn
\ee
where $u=1,\cdots, 4n_h$.
}
\be
k_1=\partial_a \quad \textrm{and} \quad k_{2}= \zeta \partial_{\tilde{\zeta}}-\tilde{\zeta }\partial_{\zeta},
\ee
and the embedding tensors 
\begin{align}
 \Theta_0^{\phantom{0}1}=1, \quad \Theta^{01}=-c, 
 \quad \Theta_n^{\phantom{n}2}=1, \quad \Theta^{n2}=0,
\end{align}
where $n=1,2,3$.
Among the isometries of the quaternionic manifold, two isometries are chosen to gauge the theory. 
The first Killing vector $k_1$ generates SO(1,1), which is gauged dyonically. The second Killing vector generates electrically gauged U(1). Although there are three vector fields, only the sum of them participate in gauging.
The metric of the quaternionic manifold can be easily read off from \eqref{Lscalar} and is
\be
h_{uv}dq^{u}dq^{v}=
  (\partial_\mu \phi)^2 
+\dfrac{1}{4} e^{4\phi} \left( \partial_\mu a +\dfrac{1}{2}
 \left(\zeta \partial_\mu \tilde{\zeta}- \tilde{\zeta} \partial_\mu \zeta  \right)\right)^2
+\dfrac{1}{4} e^{2\phi} \left((\partial_\mu \zeta)^2 +(\partial_\mu \tilde{\zeta})^2 \right).
\ee
To calculate the moment maps, we take the orthonormal frame as
\be
E^1= d\phi, \quad E^2= \dfrac{1}{2}e^{2\phi}\left(da+\dfrac{1}{2}\left(\zeta d\tilde{\zeta}-\tilde{\zeta} d\zeta\right)\right),\quad
E^3=\dfrac{1}{2}e^{\phi}d\zeta, \quad E^4=\dfrac{1}{2}e^{\phi}d\tilde{\zeta}.
\ee
and construct an SU(2) Lie-algebra valued two-form as
\begin{align}
J^1&= 2 \left(E^1 \wedge E^3 + E^2 \wedge E^4 \right),\nn\\
J^2&= 2 \left(E^1 \wedge E^4 + E^3 \wedge E^2 \right),\nn\\
J^3&= 2 \left(E^1 \wedge E^2 + E^4 \wedge E^3 \right),
\end{align}
which satisfies the quaternionic algebra
\be
J^x J^y = - \delta^{xy}+\epsilon^{xyz}J^z.
\ee
This quaternionic K\"{a}hler form is covariantly closed with respect to the SU(2) connection one-form $w^x$
\be
dJ^x +\epsilon^{xyz}w^y \wedge J^z=0,
\ee
and proportional to the SU(2) curvature
\be
dw^x+\dfrac{1}{2}\epsilon^{xyz}w^y \wedge w^z=-J^x.
\ee
We calculate the SU(2) connection one-form $w^x$ as
\begin{align}
w^1=-e^\phi d\zeta,
\quad
w^2=-e^\phi d\tilde{\zeta},
\quad
w^3= -\dfrac{1}{4}e^{2\phi}\left( 2da-\tilde{\zeta}d\zeta+\zeta d\tilde{\zeta} \right).
\end{align}
Finally, we obtain the moment maps as
\begin{align}
P_1^{\phantom{1}x}&=\left(0,\,0,\,-\dfrac{1}{2}\,e^{2\phi}\right),\nn\\
P_{2}^{\phantom{2}x}&= \left(e^{\phi}\,\tilde{\zeta},\, -e^{\phi}\,\zeta,\,
1-\dfrac{1}{4}(\zeta^2+\tilde{\zeta}^2)\,e^{2\phi}\right),
\end{align}
which satisfy 
\be
k_\alpha^{\phantom \alpha u}\, J^x_{uv}\, dq^v =\dfrac{1}{2}\left(dP_\alpha^{\phantom \alpha x}+\epsilon^{xyz}\,
 w^y\, P_\alpha^{\phantom \alpha z}\right).
\ee

Using all the results we obtained in this section, the canonical expression for the ${\mathcal{N}}=2$ scalar potential (See, for example, eq. 3.39 of \cite{Guarino:2015qaa}.)
\be
V= 4\, g^2\, \Theta_M^{\phantom{M}\alpha}\Theta_N^{\phantom{N}\beta} \Big[
4\, e^K\, X^M \bar{X}^N h_{uv}\, k^u_{\phantom{u}\alpha} k^v_{\phantom{v}\beta}
+P_\alpha^{\phantom{\alpha}x} P_\beta^{\phantom{\beta}x}
\left(K^{t_i \bar{t}_i}f_{t_i}^{\phantom{t_i}M} \bar{f}_{\bar{t}_i}^{\phantom{t_i}N}
-3\,e^K\, X^M \bar{X}^N\right)\Big],
\ee
reproduces exactly the same scalar potential \eqref{pot}.
The last term of the above scalar potential can be rewritten in terms of the superpotential as
\be
 g^2\, \Theta_M^{\phantom{M}\alpha}\Theta_N^{\phantom{N}\beta}
 P_\alpha^{\phantom{\alpha}x} P_\beta^{\phantom{\beta}x}
 e^K\, X^M \bar{X}^N
 =|\mathcal{W}_+|^2+|\mathcal{W}_-|^2.
\ee
Similar analysis for the electrically gauged theory can be found in \cite{Bobev:2010ib}. (See eq. B.42.)
\subsection{An ${\cal{N}}=1$ formulation}
Since the gauge fields play no role in the domain wall solutions,  let us focus on the neutral scalar fields in this section. We truncate away the vector fields and the charged scalar fields $(a, \, \beta)$. Then, we write the action with eight neutral real scalars and the metric, which can be re-organized into the bosonic action of ${\cal{N}}=1$ supergravity coupled to four chiral multiplets in ${\cal{N}}=1$ formulation. The bosonic action becomes
\begin{align}\label{Laction}
S=&\dfrac{1}{16 \pi G_4} \int d^4 x \sqrt{-g}
\Big[R-\dfrac{1}{2} \sum_{n=1}^3 \left((\partial_\mu \varphi_n)^2 + e^{2\varphi_n}(\partial_\mu \chi_n)^2\right)-2 \left((\partial_\mu \phi)^2+e^{2\phi_n}(\partial_\mu \rho)^2\right)
-V\Big], \nn\\
=&\dfrac{1}{8 \pi G_4} \int d^4 x \sqrt{-g} 
\Big[\dfrac{1}{2}R-\mathcal{K}_{\alpha \bar{\beta}}\, \partial_\mu z^\alpha \, \partial^\mu \bar{z}^{\bar{\beta}}-\dfrac{1}{2}V
\Big],
\end{align}
where we identify the K\"{a}hler potential 
\be
\mathcal{K} = - \textrm {log} (1- z_1 \bar{z}_1) (1- z_2 \bar{z}_2) (1- z_3 \bar{z}_3) (1- \zeta_{12} \bar{\zeta}_{12})^4,
\ee
and the K\"{a}hler metric
\begin{align}
\mathcal{K}_{\alpha \bar{\beta}} &\equiv \dfrac{\partial^2 \mathcal{K}}{\partial z^{\alpha} \partial \bar{z}^{\bar\beta}},\nn\\
&= \textrm{diag} \left(\dfrac{1}{(1-z_1 \bar{z}_1)^2},\dfrac{1}{(1-z_2 \bar{z}_2)^2},\dfrac{1}{(1-z_3 \bar{z}_3)^2},\dfrac{4}{(1-\zeta_{12} \bar{\zeta}_{12})^2} \right).
\end{align}
Here we introduce a new chiral multiplet scalar $z^4 \equiv \zeta_{12}$, which is a neutral complex scalar in the ${\cal{N}}=2$ hypermultiplet. Then the scalar potential \eqref{SP} can be written in terms of the superpotential and the inverse K\"{a}hler metric as
\footnote{We follow the convention of \cite{Freedman:2013ryh}. See sec. 3.2.}
\be
\dfrac{1}{2}V= 16\, \mathcal{K}^{\alpha\bar{\beta}}\, \partial_{\alpha} W\, \partial_{\bar{\beta}} W -12 W^2.
\ee
In ${\mathcal N}=1$ supergravity, the superpotential can be expressed as
\be
V_{F}=e^{\mathcal{K}} \left(-3\, \mathcal{V}\, \bar{\mathcal{V}} 
+ \mathcal{K}^{\alpha\bar{\beta}}\, \nabla_\alpha \mathcal{V}\, \nabla_{\bar{\beta}}\bar{\mathcal{V}} \right).
\ee
where the K\"{a}hler covariant derivative is
\be
\nabla_\alpha \mathcal{V} \equiv \partial_\alpha \mathcal{V} + \partial_\alpha \mathcal{K}\, \mathcal{V}.
\ee
By equating the two expressions for the scalar potential, we identify the holomorphic superpotential $\mathcal{V}$ as
\be
\mathcal{V}= 2\, e^{-\mathcal{K}/2}\, \mathcal{W}.
\ee
Here, we used the following identities 
\be
\partial_\alpha \mathcal{K} -2\, \partial_\alpha\, \textrm{log}\bar{\mathcal{W}}=0,
\quad \textrm{and} \quad
\partial_{\bar\alpha} \mathcal{K} -2\, \partial_{\bar\alpha}\, \textrm{log}\mathcal{W}=0,
\ee
and 
\be
\nabla_\alpha \mathcal{V}
=4\, e^{-\mathcal{K}/2}\,\dfrac{W}{ \bar{\mathcal{W}}}\, \partial_\alpha W,
\quad \textrm{and} \quad
\nabla_{\bar\alpha} \mathcal{\bar{V}}
=4\, e^{-\mathcal{K}/2}\,\dfrac{W}{ \mathcal{W}}\, \partial_{\bar\alpha}\, W,
\ee
 which can be obtained by using the holomorphic property of the superpotential $\mathcal{V}$. Using the K\"{a}hler potential $\mathcal{K}$ and the holomorphic superpotential $ \mathcal{V}$, one can easily write down the supersymmetry variations of the gravitino and the dilatino. 

At this point, it is worthwhile to note that the holomorphic superpotential can be written in a simpler way.
Under a K\"{a}hler transformation
\be
\mathcal{K} \longrightarrow \mathcal{\tilde K}(z,\bar{z})=\mathcal{K}(z,\bar{z})+f(z)+\bar{f}(\bar{z}),
\ee
a K\"{a}hler metric $\mathcal{K}_{\alpha \bar{\beta}}$ is invariant
while the holomorphic superpotential $\mathcal{V}$ transforms as
\be
\mathcal{V}(z) \longrightarrow \mathcal{\tilde V}(z)=e^{-f(z)}\, \mathcal{V}(z).
\ee
Now, let us consider a K\"{a}hler transformation with
\be
f=-\textrm{log} \left(i \dfrac{4}{(1-z_1)(1-z_2)(1-z_3)(1-\zeta_{12})^4 }\right).
\ee
Then, we obtain the K\"{a}hler potential 
\be
\mathcal{\tilde K} = - \textrm {log} \dfrac{16(1- z_1 \bar{z}_1) (1- z_2 \bar{z}_2) (1- z_3 \bar{z}_3) (1- \zeta_{12} \bar{\zeta}_{12})^4}{(1-z_1)(1-z_2)(1-z_3)(1-\zeta_{12})^4 
(1-\bar{z}_1)(1-\bar{z}_2)(1-\bar{z}_3)(1-\bar\zeta_{12})^4}.
\ee
The form of the holomorphic superpotential is simplified in terms of the complex fields $(t_n, \,u)$ as
\begin{align}
\mathcal{\tilde V}
&=m-i\, g \dfrac{(1+z_1)(1+z_2)(1+z_3)}{(1-z_1)(1-z_2)(1-z_3)}
-2i\, g\left(\dfrac{1+z_1}{1-z_1}+\dfrac{1+z_2}{1-z_2}+\dfrac{1+z_3}{1-z_3} \right) \dfrac{(1+\zeta_{12})^2}{(1-\zeta_{12})^2}\\
&= m+g\, t_1 t_2 t_3 +2g(t_1+t_2+t_3)u^2.
\end{align}
Note that $W^2=\mathcal{W} \bar{\mathcal{W}}=\dfrac{1}{4}\mathcal{V} \bar{\mathcal{V}} e^{\mathcal{K}}$ is invariant under a K\"{a}hler transformation.

\subsection{Critical points}
As we already discussed in the previous section, the scalar potential \eqref{pot} of U(1)$^2$-invariant sector of ${\cal N}=8 $ dyonic ISO(7) gauged supergravity theory reduceㄴ to that of the SU(3)-invariant theory when $ \varphi_1=\varphi_2=\varphi_3$ and $\chi_1=\chi_2=\chi_3$. Hence, the U(1)$^2$ truncated theory
includes all the critical points which exist in the SU(3)-invariant sector. There are three supersymmetric fixed points with ${\cal N}=2$ supersymmetry SU(3)$\times$U(1) bosonic symmetry,\, ${\cal N}=1 $ G$_2$ and ${\cal N}=1$ SU(3). Additionally there are five non-supersymmetric fixed points. See Table 3 in \cite{Guarino:2015qaa} and Table 1 in \cite{Guarino:2016ynd} for the details. 

Here, we are interested in the ${\cal N}=2$ SU(3)$\times$U(1) fixed point located at
\be
z_{1*}=z_{2*}=z_{3*}=\dfrac{-2+(\sqrt{3}-i)c^{1/3}}{2+(\sqrt{3}-i)c^{1/3}},
\quad \quad 
\zeta_{12*}=\dfrac{-2+\sqrt{2}c^{1/3}}{2+\sqrt{2}c^{1/3}}.
\ee
One can compute the mass spectrum of the scalar fields and the corresponding conformal dimensions
 of the dual operators around this AdS fixed point. They are summarized in Table \ref{massU(1)2}.
\begin{table}
\begin{center}
\begin{tabular}{ c || c c c c || c c c c }
 $M^2 {L_*}^2$ & $3-\sqrt{17}$ & $2$ & $2$ & $3+\sqrt{17}$ &
 $-2$ &$-2$ &$-2$ &$-2$ \\ \hline \hline
$\Delta_+$ & $\frac{1+\sqrt{17}}{2}$ & $\frac{3+\sqrt{17}}{2}$       &$\frac{3+\sqrt{17}}{2}$ &$\frac{5+\sqrt{17}}{2}$
&$2$ &$2$ &$2$ &$2$  \\ \medskip
$\Delta_-$ & $\frac{5-\sqrt{17}}{2}$ & $\frac{3-\sqrt{17}}{2}$      &$\frac{3-\sqrt{17}}{2}$ &$\frac{1-\sqrt{17}}{2}$
&$1$ &$1$ &$1$ &$1$  \\ 
\end{tabular}
\caption{Scalar mass spectrums and conformal dimensions of dual operators around ${\cal N}=2$ SU(3)$\times$U(1) AdS fixed point} \label{massU(1)2}
\end{center}
\end{table}
The masses of the first four modes coincide with those of scalar fields in the SU(3)-invariant sector. The last four modes which are introduced in the U(1)$^2$ invariant truncation correspond to the tachyonic scalar fields with mass $M^2 {L_*}^2=-2$. On the dual field theory side, we expect that there are five relevant operators, one with dimension $\frac{1+\sqrt{17}}{2}$ and four with dimension 2. We emphasize that the last four relevant operators are introduced as a result of R-charge deformation of the dual field theory \eqref{FE} and induce RG-flows from the UV fixed point.
\section{Discussions}
We revisited the U(1)$^2$-invariant sector of $D=4$, ${\mathcal N=8}$ dyonic  ISO(7) gauged supergravity, which was recently constructed in \cite{Guarino:2017pkw} and studied further in \cite{Azzurli:2017kxo, Hosseini:2017fjo, Benini:2017oxt}. 
We focused on the aspects of the U(1)$^2$-invariant sector as the gravity duals of R-charge deformations of Guarino, Jafferis, Varela's theory \cite{Guarino:2015jca, Fluder:2015eoa}.  On the field theory side, the general assignments of R-charges  correspond to introducing mass terms, which induce RG-flows. 
As a first step towards understanding this RG-flows in a holographic setup, we have constructed the U(1)$^2$ truncated theory using the embedding tensor formalism.

In this paper, we presented explicit forms of the superpotential. It enables us to calculate the mass spectra of the scalar fields around an ${\mathcal N=2}$ fixed point. As a result of the U(1)$^2$ truncation, we obtained the four additional tachyonic scalar fields which can induce RG-flows from the AdS fixed point. They correspond to the bosonic and fermionc mass operators in the dual field theory. We also rewrote the bosonic action in an ${\mathcal N=1}$ language and identified the holomorphic superpotential and the K\"{a}hler  potential. It will allows us to obtain a set of BPS equations by requiring the vanishing of  the fermionic supersymmetry variation.
 
We are eventually interested in the field theories defined on a sphere to use the localization technique and compute the exact quantity such as the partition function, when there are mass deformations as well as the conformal point. Hence, the boundary of dual gravity solutions we are looking for should be a sphere on which the field theories are defined.
To be more specific, the holographic RG-flows should be described by the sphere-sliced domain wall, and not by the flat domain wall \cite{Freedman:2013ryh, Bobev:2013cja, Bobev:2016nua}. With as many as four complex scalar fields, 
solving the BPS equations is rather complicated mission. We aim to deal with numerical solutions to supergravity BPS equations, and reproduce the field theoretic calculation of the partition function in the near future \cite{kks}.


\begin{acknowledgments}
This manuscript is based on talks given by the authors at various conferences, including ``Workshop on Fields, Strings and Gravity" at KIAS in February 2017 (HK), Korean physical society meeting in April 2017 (MS), APCTP focus program ``Geometry and Holography for Quantum Criticality" in Pohang, August 2017 (HK), ``International workshop for string theory and cosmology", in Busan, August 2017 (HK), and ``East Asia Joint Workshop on Fields and Strings", at KEK Japan, November 2017 (NK). We thank the audience there for comments and discussions. 
This work was supported by a research grant from Kyung Hee University in 2016 (KHU-20160698). 
\end{acknowledgments}

\appendix
\section{useful formulas}\label{append}
In this appendix, we collect the useful formulas needed to construct D=4 ${\cal N}=8$ dyonic ISO(7) gauged supergravity in the main text. The details can be found, for example, in \cite{Guarino:2015qaa}.
The explicit form of the E$_{7(7)}$ generators is
\footnote{In this appendix, $A, B =1, \cdots, 8$ is a fundamental index of SL(8).}
\begin{align}\label{gen}
&[t_A^{\phantom A B}]_{[CD]}^{\phantom{[CD]}[EF]} 
=4 \left(\delta_{[C}^B \delta_{D]A}^{EF}+\dfrac{1}{8}\delta_{A}^B \delta_{CD}^{EF}\right)
\quad
&\textrm{and}
\quad
&[t_A^{\phantom A B}]^{[EF]}_{\phantom{[CD]}[CD]} =-[t_A^{\phantom A B}]_{[CD]}^{\phantom{[CD]}[EF]},\nn\\
&[t_{ABCD}]_{[EF][GH]}=\dfrac{2}{4!}\epsilon_{ABCDEFGH}
\quad
&\textrm{and}
\quad
&[t_{ABCD}]^{[EF][GH]}=2\delta_{ABCD}^{EFGH}.
\end{align}
We can construct the X-tensors using the embedding tensors and the generators as
\be\label{xtensor}
X_{\mathbb{MN}}^{\phantom{\mathbb{MN}}\mathbb{P}}
=\Theta_{\mathbb{M}}^{\phantom M \alpha} [t_\alpha]_{\mathbb{N}}^{\phantom{\mathbb{N}}\mathbb{P}}
=\Theta_{\mathbb{M}\phantom {C}D}^{\phantom M C} [t_C^{\phantom C D}]_{\mathbb{N}}^{\phantom{\mathbb{N}}\mathbb{P}}.
\ee
The elements of X-tensors are given by
\begin{align}\label{xtensorform}
&X_{[AB][CD]}^{\phantom{[AB][CD]} [EF]}=-X_{[AB]\phantom{[AB]}[CD]}^{\phantom{[AB]} [EF]}
=-8\, \delta_{[A}^{[E}\theta_{B][C} \delta_{D]}^{F]},\nn\\
&X^{[AB]\phantom{[AB]}[EF]}_{\phantom{[AB]} [CD]}=-X^{[AB][EF]}_{\phantom{[AB][CD]} [CD]}
=-8\, \delta_{[C}^{[A}\xi^{B][E} \delta_{D]}^{F]}.
\end{align}

\bibliography{ref}{}

\providecommand{\href}[2]{#2}\begingroup\raggedright\begin{thebibliography}{10}

\bibitem{Maldacena:1997re}
J.~M. Maldacena, {\it {The Large N limit of superconformal field theories and
  supergravity}},  {\em Int. J. Theor. Phys.} {\bf 38} (1999) 1113--1133,
  [\href{http://arxiv.org/abs/hep-th/9711200}{{\tt hep-th/9711200}}]. [Adv.
  Theor. Math. Phys.2,231(1998)].

\bibitem{Aharony:2008ug}
O.~Aharony, O.~Bergman, D.~L. Jafferis, and J.~Maldacena, {\it {N=6
  superconformal Chern-Simons-matter theories, M2-branes and their gravity
  duals}},  {\em JHEP} {\bf 10} (2008) 091,
  [\href{http://arxiv.org/abs/0806.1218}{{\tt arXiv:0806.1218}}].

\bibitem{DallAgata:2011aa}
G.~Dall'Agata and G.~Inverso, {\it {On the Vacua of N = 8 Gauged Supergravity
  in 4 Dimensions}},  {\em Nucl. Phys.} {\bf B859} (2012) 70--95,
  [\href{http://arxiv.org/abs/1112.3345}{{\tt arXiv:1112.3345}}].

\bibitem{DallAgata:2012mfj}
G.~Dall'Agata, G.~Inverso, and M.~Trigiante, {\it {Evidence for a family of
  SO(8) gauged supergravity theories}},  {\em Phys. Rev. Lett.} {\bf 109}
  (2012) 201301, [\href{http://arxiv.org/abs/1209.0760}{{\tt
  arXiv:1209.0760}}].

\bibitem{DallAgata:2014tph}
G.~Dall'Agata, G.~Inverso, and A.~Marrani, {\it {Symplectic Deformations of
  Gauged Maximal Supergravity}},  {\em JHEP} {\bf 07} (2014) 133,
  [\href{http://arxiv.org/abs/1405.2437}{{\tt arXiv:1405.2437}}].

\bibitem{Guarino:2015jca}
A.~Guarino, D.~L. Jafferis, and O.~Varela, {\it {String Theory Origin of Dyonic
  N=8 Supergravity and Its Chern-Simons Duals}},  {\em Phys. Rev. Lett.} {\bf
  115} (2015), no.~9 091601, [\href{http://arxiv.org/abs/1504.08009}{{\tt
  arXiv:1504.08009}}].

\bibitem{Guarino:2015qaa}
A.~Guarino and O.~Varela, {\it {Dyonic ISO(7) supergravity and the duality
  hierarchy}},  {\em JHEP} {\bf 02} (2016) 079,
  [\href{http://arxiv.org/abs/1508.04432}{{\tt arXiv:1508.04432}}].

\bibitem{Guarino:2015vca}
A.~Guarino and O.~Varela, {\it {Consistent $ \mathcal{N}=8 $ truncation of
  massive IIA on S$^{6}$}},  {\em JHEP} {\bf 12} (2015) 020,
  [\href{http://arxiv.org/abs/1509.02526}{{\tt arXiv:1509.02526}}].

\bibitem{Guarino:2016ynd}
A.~Guarino, J.~Tarrio, and O.~Varela, {\it {Romans-mass-driven flows on the
  D2-brane}},  {\em JHEP} {\bf 08} (2016) 168,
  [\href{http://arxiv.org/abs/1605.09254}{{\tt arXiv:1605.09254}}].

\bibitem{Guarino:2017eag}
A.~Guarino and J.~Tarr{\'\i}o, {\it {BPS black holes from massive IIA on
  S$^{6}$}},  {\em JHEP} {\bf 09} (2017) 141,
  [\href{http://arxiv.org/abs/1703.10833}{{\tt arXiv:1703.10833}}].

\bibitem{Pang:2017omp}
Y.~Pang, J.~Rong, and O.~Varela, {\it {Spectrum universality properties of
  holographic Chern-Simons theories}},
  \href{http://arxiv.org/abs/1711.07781}{{\tt arXiv:1711.07781}}.

\bibitem{Araujo:2016jlx}
T.~R. Araujo and H.~Nastase, {\it {Observables in the
  Guarino-Jafferis-Varela/CS-SYM duality}},  {\em JHEP} {\bf 07} (2017) 020,
  [\href{http://arxiv.org/abs/1609.08008}{{\tt arXiv:1609.08008}}].

\bibitem{Araujo:2017hvi}
T.~Araujo, G.~Itsios, H.~Nastase, and E.~{\'O}. Colg{\'a}in, {\it {Penrose
  limits and spin chains in the GJV/CS-SYM duality}},  {\em JHEP} {\bf 12}
  (2017) 137, [\href{http://arxiv.org/abs/1706.02711}{{\tt arXiv:1706.02711}}].

\bibitem{Fluder:2015eoa}
M.~Fluder and J.~Sparks, {\it {D2-brane Chern-Simons theories: F-maximization =
  a-maximization}},  {\em JHEP} {\bf 01} (2016) 048,
  [\href{http://arxiv.org/abs/1507.05817}{{\tt arXiv:1507.05817}}].

\bibitem{Jafferis:2010un}
D.~L. Jafferis, {\it {The Exact Superconformal R-Symmetry Extremizes Z}},  {\em
  JHEP} {\bf 05} (2012) 159, [\href{http://arxiv.org/abs/1012.3210}{{\tt
  arXiv:1012.3210}}].

\bibitem{Jafferis:2011zi}
D.~L. Jafferis, I.~R. Klebanov, S.~S. Pufu, and B.~R. Safdi, {\it {Towards the
  F-Theorem: N=2 Field Theories on the Three-Sphere}},  {\em JHEP} {\bf 06}
  (2011) 102, [\href{http://arxiv.org/abs/1103.1181}{{\tt arXiv:1103.1181}}].

\bibitem{Pufu:2016zxm}
S.~S. Pufu, {\it {The F-Theorem and F-Maximization}},  2016.
\newblock \href{http://arxiv.org/abs/1608.02960}{{\tt arXiv:1608.02960}}.

\bibitem{Freedman:2013ryh}
D.~Z. Freedman and S.~S. Pufu, {\it {The holography of $F$-maximization}},
  {\em JHEP} {\bf 03} (2014) 135, [\href{http://arxiv.org/abs/1302.7310}{{\tt
  arXiv:1302.7310}}].

\bibitem{Bobev:2013cja}
N.~Bobev, H.~Elvang, D.~Z. Freedman, and S.~S. Pufu, {\it {Holography for $N =
  2^*$ on $S^4$}},  {\em JHEP} {\bf 07} (2014) 001,
  [\href{http://arxiv.org/abs/1311.1508}{{\tt arXiv:1311.1508}}].

\bibitem{Bobev:2016nua}
N.~Bobev, H.~Elvang, U.~Kol, T.~Olson, and S.~S. Pufu, {\it {Holography for $
  \mathcal{N} $ = 1$^{*}$ on S$^{4}$}},  {\em JHEP} {\bf 10} (2016) 095,
  [\href{http://arxiv.org/abs/1605.00656}{{\tt arXiv:1605.00656}}].

\bibitem{Guarino:2017pkw}
A.~Guarino, {\it {BPS black hole horizons from massive IIA}},  {\em JHEP} {\bf
  08} (2017) 100, [\href{http://arxiv.org/abs/1706.01823}{{\tt
  arXiv:1706.01823}}].

\bibitem{Hosseini:2017fjo}
S.~M. Hosseini, K.~Hristov, and A.~Passias, {\it {Holographic microstate
  counting for AdS$_{4}$ black holes in massive IIA supergravity}},  {\em JHEP}
  {\bf 10} (2017) 190, [\href{http://arxiv.org/abs/1707.06884}{{\tt
  arXiv:1707.06884}}].

\bibitem{Azzurli:2017kxo}
F.~Azzurli, N.~Bobev, P.~M. Crichigno, V.~S. Min, and A.~Zaffaroni, {\it {A
  Universal Counting of Black Hole Microstates in AdS$_4$}},
  \href{http://arxiv.org/abs/1707.04257}{{\tt arXiv:1707.04257}}.

\bibitem{Benini:2017oxt}
F.~Benini, H.~Khachatryan, and P.~Milan, {\it {Black hole entropy in massive
  Type IIA}},  \href{http://arxiv.org/abs/1707.06886}{{\tt arXiv:1707.06886}}.

\bibitem{kks}
H.~Kim, N.~Kim, and M.~Suh, {\it work in progress}.

\bibitem{Guarino:2017jly}
A.~Guarino, {\it {Hypermultiplet gaugings and supersymmetric solutions from 11D
  and massive IIA supergravity on H$^{(p,q)}$ spaces}},
  \href{http://arxiv.org/abs/1712.09549}{{\tt arXiv:1712.09549}}.

\bibitem{Duff:1986hr}
M.~J. Duff, B.~E.~W. Nilsson, and C.~N. Pope, {\it {Kaluza-Klein
  Supergravity}},  {\em Phys. Rept.} {\bf 130} (1986) 1--142.

\bibitem{Bobev:2010ib}
N.~Bobev, N.~Halmagyi, K.~Pilch, and N.~P. Warner, {\it {Supergravity
  Instabilities of Non-Supersymmetric Quantum Critical Points}},  {\em Class.
  Quant. Grav.} {\bf 27} (2010) 235013,
  [\href{http://arxiv.org/abs/1006.2546}{{\tt arXiv:1006.2546}}].

\bibitem{Nicolai:1985hs}
H.~Nicolai and N.~P. Warner, {\it {The SU(3) X U(1) Invariant Breaking of
  Gauged $N=8$ Supergravity}},  {\em Nucl. Phys.} {\bf B259} (1985) 412.

\bibitem{Ahn:2008ya}
C.~Ahn, {\it {Holographic Supergravity Dual to Three Dimensional N=2 Gauge
  Theory}},  {\em JHEP} {\bf 08} (2008) 083,
  [\href{http://arxiv.org/abs/0806.1420}{{\tt arXiv:0806.1420}}].

\bibitem{Klebanov:2008vq}
I.~Klebanov, T.~Klose, and A.~Murugan, {\it {AdS(4)/CFT(3) Squashed, Stretched
  and Warped}},  {\em JHEP} {\bf 03} (2009) 140,
  [\href{http://arxiv.org/abs/0809.3773}{{\tt arXiv:0809.3773}}].

\bibitem{deWit:2007kvg}
B.~de~Wit, H.~Samtleben, and M.~Trigiante, {\it {The Maximal D=4
  supergravities}},  {\em JHEP} {\bf 06} (2007) 049,
  [\href{http://arxiv.org/abs/0705.2101}{{\tt arXiv:0705.2101}}].

\bibitem{Andrianopoli:1996cm}
L.~Andrianopoli, M.~Bertolini, A.~Ceresole, R.~D'Auria, S.~Ferrara, P.~Fre, and
  T.~Magri, {\it {N=2 supergravity and N=2 superYang-Mills theory on general
  scalar manifolds: Symplectic covariance, gaugings and the momentum map}},
  {\em J. Geom. Phys.} {\bf 23} (1997) 111--189,
  [\href{http://arxiv.org/abs/hep-th/9605032}{{\tt hep-th/9605032}}].

\bibitem{Trigiante:2016mnt}
M.~Trigiante, {\it {Gauged Supergravities}},  {\em Phys. Rept.} {\bf 680}
  (2017) 1--175, [\href{http://arxiv.org/abs/1609.09745}{{\tt
  arXiv:1609.09745}}].

\end{thebibliography}\endgroup
\end{document}